\documentclass[
]{ceurart}

\usepackage{listings}
\usepackage{todonotes}

\begin{document}

\copyrightyear{2026}
\copyrightclause{Copyright for this paper by its authors.
  Use permitted under Creative Commons License Attribution 4.0
  International (CC BY 4.0).}

\conference{Joint Proceedings of REFSQ-2026 Workshops, Doctoral Symposium, Posters \& Tools Track, and Education and Training Track. Co-located with REFSQ 2026. Poznan, Poland, March 23-26, 2026}

\title{Towards Viewpoint-centric Artifact-based Regulatory Requirements Engineering for Compliance by Design}

\author[1.2]{Oleksandr Kosenkov}[
orcid=0000-0002-9971-1130,
email=oleksandr.kosenkov@bth.se,
url=https://regulatory-re.com/
]

\address[1]{Blekinge Institute of Technology,
            Valhallavägen 10, 371 79 Karlskrona, Sweden}
\address[2]{fortiss GmbH,
            Guerickestr. 25, 80805, Munich, Germany}
\maketitle

\begin{abstract}
Processing regulations and resulting requirements to achieve regulatory compliance in software engineering (SE) is a developing challenge due to the continuously growing amount, complexity, and expanding scope of regulations. Despite the growing amount of newly suggested regulatory requirements engineering (RE) approaches by the research community, industry remains under pressure to assure their integration into their RE and overall software development life cycle (SDLC) practices to facilitate a seamless and legally valid compliance by design.
As of today, we still have limited empirical understanding of how this can be achieved. 
Such integration should avoid additional burdens and address the demands of legal knowledge intensity, cross-functional communication and consistency between different involved viewpoints. Intermediary results of this doctoral study showed that regulatory RE has peculiarities distinguishing it from the engineering of other requirements. Oftentimes, organizations establish standalone regulatory RE processes on the organizational level. However, software development teams usually approach compliance by design in an ad-hoc manner, rather than in a systematic way. Among other, because of the complexity of the coordination between the involved viewpoints.
The goal of this paper is to report and get feedback about the synthesis and future evaluation of our Artefact Model for Regulatory Requirements Engineering (AM4RRE) for a integrated compliance by design. We hope this paper will spark discussions about regulatory RE and help us refine plans for the final stage of the doctoral study.
\end{abstract}

\section{Introduction}
Regulatory requirements engineering (RE)---the processing of regulations as a source of requirements for implementing regulatory compliance in software engineering (SE)---has been in scope of research activities for two decades now~\cite{syed2010emerging}. Since then, the regulations have evolved and became more focused on regulating software-intensive systems and services. Among other, a principle of compliance by design was introduced to emphasize that compliance controls should be implemented throughout the SDLC and in software design specifically. In response to the growing number of regulations, the RE research community has proposed a significant number of new methods and tools. However, the software industry mainly continues to cope with the regulatory compliance using isolated methods, staying under the pressure of finding more effective and systematic approaches~\cite{klymenko2022understanding}.

In practice, regulatory RE cannot be implemented by simply involving legal experts as these are often unavailable, as it cannot be simply implemented with one or two additional RE techniques out of the box. Regulatory RE has regulations as a very specific source of requirements and is driven by the purpose of compliance implementation and not typical business goals. Hence, it requires seamless integration with existing RE methods and broader SDLC processes through corresponding engineering artefacts. To develop a regulatory RE approach for a seamless compliance by design, we established an empirically sound understanding of regulatory RE in practice and challenges to it. On this basis, we developed a fundamental, conceptual model interconnecting the elementary legal domain concepts and artefacts for specification of legal viewpoint in regulatory RE and developed an approach for integrating it with engineering viewpoints. In this paper, we report the results of our research, the initial version of our synthesized regulatory RE approach, and our plans for its evaluation to get community feedback.

\section{Integrating Regulatory RE into RE and SDLC Practices }\label{sec:challenge}
Next, we outline five characteristics that distinguish regulatory RE from other RE practices and constitute the challenge of integration of regulatory aspects into RE and more generally into SDLC practices.

First, regulatory RE and the ultimate goal of verifiable regulatory compliance of software systems require consistency, completeness and correctness in the RE results usually coming in the form of artefacts. This is hardly achievable using the widespread process-based RE approaches due to significant variability and dynamics in real-world projects. Further, nowadays, software development is usually based on agile methodology favoring ad-hoc approaches and often spontaneous human interaction. Despite the claims that agile methods support the implementation of regulatory compliance ``out of the box'', we found out that in practice, companies are still struggling to integrate compliance into scaled agile frameworks~\cite{kosenkov2024regulatory}. Herewith, it is challenging to establish an RE process supporting consistency, completeness and correctness in dynamic agile development.

Second, regulatory requirements are a ``natural afterthought'' as they derive from the interpretation of regulations in relation to a software system or corresponding requirements. Due to this, regulatory requirements are usually engineered not simultaneously with other requirements, but when at least initial versions of requirements and/or system specification artefacts are available.

Third, regulatory requirements have impact on other non-regulatory requirements and may demand the revision of the latter. As a result, iterations between regulatory and non-regulatory requirements can be required. We found out that in practice, companies do in fact establish a standalone regulatory RE process executed on top of their usual RE and SDLC processes~\cite{kosenkov2024regulatory}. As a result, regulatory RE needs to be appropriately integrated with the engineering of non-regulatory requirements.

Fourth, regulations often contain both problem space and solution space-related demands which need to be considered both in requirements and system specifications completed and maintained by different roles~\cite{kosenkov2025privacy}. As a result, regulatory RE methods should be integrated into SDLC and support seamless communication of regulatory demands into software architectures. Consistency should be further preserved between requirements and software architecture viewpoints to make sure iterative refinement and decisionmaking about regulatory requirements are possible.

Fifth, interaction between different viewpoints in regulatory RE is essential~\cite{kosenkov2024regulatory} and hardly fully automatable. Regulatory RE requires viewpoint-specific in-depth domain knowledge~\cite{kosenkov2025systematic} (which is hard to explicate~\cite{kosenkov2024regulatory}), and achievement of viewpoint-specific goals ~\cite{kosenkov2026towards}. In the case of regulatory RE, engineering-legal interaction is challenging~\cite{bednar2019engineering}, because of conflicting goals and complexity of translation between the viewpoints caused by the differences in legal and SE viewpoints.

Taking into account these challenges, and our intermediary results achieved so far throughout the PhD studies, this thesis suggests that regulatory RE can be implemented through coordination of the involved viewpoints---communication between different viewpoints for the achievement of a common goal of compliance in regulatory RE. Such coordination should (1) support the presentation of each viewpoint in a way capturing relevant knowledge and goals in regulatory compliance (or if necessary compensate for the absence of viewpoint-specific knowledge), and (2) support flexible communication between the roles in each viewpoint. Such coordination is hard to establish and manage using activity-oriented approaches, which does inter alia not allow to effectively account for completeness, consistency, and correctness of the information specified in regulatory RE. Instead, we suggest using an artefact-based approach in which RE is guided not through a reference model dictating activities, but with one outlining the content of the artefacts that are specified and their relationships. For viewpoints coordination in regulatory RE we are synthesizing and planning to evaluate the Artefact Model for Regulatory Requirements Engineering (AM4RRE).

Research conducted in this thesis has been guided by the following three research questions:
\begin{enumerate}
\item[RQ1:] What are the challenges to regulatory RE and approaches to privacy by design in the literature?

\item[RQ2:] How is regulatory RE (including viewpoint coordination) implemented in practice and what RE methods are required to address regulatory RE for compliance by design?

\item[RQ3:] How can coordination between viewpoints contribute to regulatory RE for compliance by design?
\end{enumerate}

\section{RQ1: Analysis of the State of Reported Evidence}\label{sec:stateResearch}
For the exploration of the existing state of research, we conducted a systematic mapping study~\cite{kosenkov2025systematic}, and executed systematic literature reviews in studies for AM4RRE components synthesis or validation~\cite{kosenkov2025privacy,kosenkov2026towards}.

Among 11 categories of challenges to regulatory RE and compliance identified in our study~\cite{kosenkov2025systematic}, the top challenges were related to viewpoint-specific knowledge such as abstractness of regulations, demand for legal, IT, and privacy or security domain knowledge, and challenges in interaction between experts. Such challenges, including conflicts or changes to existing SE practices caused by regulations as well as the absence or insufficiency of existing approaches, made clear how existing RE methods are rendered insufficient for their integration into a more holistic regulatory RE as part of the overall SDLC. Another important finding was that together with software engineering roles, also legal or compliance experts, system stakeholders, or other experts were mentioned as relevant in primary studies. Many primary studies considered RE and compliance from the perspective of the whole SDLC or as a conjunction of RE and software design, quality assurance, or other SDLC stages. We found this to be aligned with the principle of compliance by design introduced in some regulations (like General Data Protection, and Cyber Resilience Act) and emphasizing the demand for consistency between RE and subsequent SDLC stages. The majority of primary studies addressed compliance with one regulation, while in practice multiple regulations need to be implemented in the same system concurrently.

In our subsequent study~\cite{kosenkov2025privacy}, we systematically reviewed the literature on methods for privacy by design. In this study, we conceptualized privacy by design as the specification of requirements and early software design in response to General Data Protection Regulation (GDPR) norms, facilitating demonstrable and verifiable compliance. Following this approach, we searched for studies explicitly addressing RE and software design and providing examples of both requirements and system (constraints) specifications. Among the five identified studies, none applied a systematic approach that also could be replicated for other regulations. We also synthesized the main characteristics of requirements engineering (RE) methods for privacy by design, namely: capturing legal domain knowledge and goals, traceability \& consistency of specifications, separation of compliance \& non-compliance concerns, system specification transparency \& overview, and specification enabling system flexibility. 

In~\cite{kosenkov2026towards}, we conducted a literature review to synthesize the goals that RE methods for privacy by design should support and to derive an assessment approach for such methods.

\section{RQ2: State of Practice Analysis}
We conducted empirical research to explore the state of practice in regulatory RE and compliance by design through five studies~\cite{klymenko2022understanding,kosenkov2024developing,kosenkov2024regulatory,kosenkov2025privacy,kosenkov2026towards}.

Our first study~\cite{klymenko2022understanding} used interview research to understand the practice used for the implementation of technical measures for GDPR compliance. It provided first evidence that regulatory RE requires profound interaction between engineering and legal or privacy-related roles. However, in many cases, there were particular go-in-between roles that intermediate such communication.

Our next study~\cite{kosenkov2024developing} approached legal experts as the roles involved in regulatory RE to better explore their viewpoint. Through focus groups with legal researchers, we identified four key challenges for regulatory requirements engineering: detachment from legal interpretation practice, the non-linear and iterative nature of legal interpretation, limited consideration of the software context, and the restricted use of legal concepts in existing regulatory RE approaches. We also found that challenges from a software engineering viewpoint can often be approached and resolved from the legal viewpoint. For example, the most frequently mentioned challenge of abstractness of regulations is in fact one of the features of regulations intended by regulators~\cite{breaux2007systematic}, and is addressed with the application of a knowledge-intensive legal interpretation process intended for concretization of abstract norms into concrete case-specific requirements~\cite{kosenkov2024developing}.

Next, we have explored the regulatory impact analysis process as one type of regulatory RE activities at the example of the European Accessibility Act compliance in large enterprises~\cite{kosenkov2024regulatory}.
We found this regulatory RE process to be executed through the coordination of multiple roles and stakeholders across different functions (cross-functional coordination) and enterprise levels (vertical coordination). Cross-functional interaction is addressed through introducing cross-functional working group and producing and tailoring artifacts. In this process, practitioners mainly experience challenges with accommodating agile development and waterfall-like compliance ways of working, implicit and undocumented knowledge, multiplicity of regulations, interpretation consistency, making analysis results applicable, demand for expert knowledge, measuring and sustaining compliance.

In~\cite{kosenkov2025privacy,kosenkov2026towards}, we have explored the importance of the five RE method characteristics for privacy by design (see above) from the perspective of practitioners involved in privacy by design implementations. The results showed that top 3 characteristics capturing legal domain knowledge, and transparency of specifications, along with traceability are the most important method characteristics. Only 2 participants out of 15 used a systematic approach (a custom developed tool and data protection impact assessment process) to privacy by design.

Finally, in~\cite{kosenkov2026towards}, we have analyzed the results of interviews with practitioners participating in privacy by design implementations and formulated 11 high-level RE-related goals that practitioners pursue in the implementation of privacy by design. On the basis of our observation that it is easier for practitioners to reason about method goals rather than method characteristics, we suggested an initial vision for a goal-centric assessment framework for privacy by design inspired by Goal Question Metric model. This assessment approach remains a subject of our ongoing work.

\section{RQ3: Synthesis of RE Model for Viewpoints Coordination}
The development of the suggested solution was reported throughout four studies~\cite{kosenkov2021vision,kosenkov2024developing,kosenkov2025privacy,kosenkov2026towards}.

In \cite{kosenkov2021vision}, we described our first abductively derived vision towards the possibility to bridge the gap between engineering and legal perspectives with artefact-based RE.

In \cite{kosenkov2024developing}, on the basis of the data collected from legal researchers, we identified the need for executing legal interpretation to concretize high-level regulatory demands into project and system-specific demands. Building on the inputs provided by legal researchers in focus groups, we identified the core legal concepts required for the analysis of regulatory texts from the legal viewpoint. On the basis of the identified legal concepts, we have developed a first vision of legal viewpoint specification composed of legal context specification and legal demands specification. We conducted a walkthrough-based validation with legal experts towards the possibility to use our developed legal viewpoint together with software context, requirements, and system specifications of AMDiRE for conducting legal interpretation independently of specific project or applicable regulations. Involved legal researchers provided overall positive feedback.

In our next study~\cite{kosenkov2025privacy}, we focused on further extending the legal viewpoint specification with the concepts required for the implementation of privacy by design according to GDPR. In this study, we have specifically focused on the validation of the process of instantiating legal concepts for GDPR and their allocation across requirements and system viewpoints as required according to the privacy by design principle. This was conducted in two phases. In the first phase, we instantiated the concepts using the whole text of GDPR and derived corresponding requirements (concept instances allocated to requirements viewpoint) and system components (concept instances allocated to system viewpoint). Next, we compared the derived requirements and system components to ones suggested in existing studies. In the second phase, we conducted a validation of the instantiation method with practitioners involved in privacy by design in practice. We provided study participants with definitions and the model of three legal concepts (incl. their relationships), examples of an annotation of selected GDPR excerpts and the resulting concept instances allocated across requirements and system levels of abstraction. We asked participants to annotate the text of other GDPR excerpts and develop corresponding models. After the exercise, participants were provided with the ground truth and asked to evaluate the provided methods according to five main method characteristics we identified in the literature review. The results suggest that the identified concepts and our approach of allocating their instances between requirements and system levels of abstraction is applicable for capturing legal domain knowledge and can be further integrated into the version of the approach we have developed previously.

In our next study~\cite{kosenkov2026towards}, we identified 11 high-level goals that RE methods for privacy by design should address on the basis of the interviews with practitioners. Taking into account the broad range of these goals, differences in goals across viewpoints, and potential variability in real world projects, we decided to incorporate the identified goal into our artefact model to facilitate coordination between viewpoints and documenting essential goals in a projects specific settings.

\section{Solution Vision}
In order to address the integration of regulatory RE into the SDLC, we extend the existing state of the art in artefact-based RE represented by the Artefact Model for Domain-independent Requirements Engineering (AMDiRE)~\cite{mendez2015artefact} into our Artefact Model for Regulatory Requirements Engineering (AM4RRE). Artifact-based RE guides RE process through artefacts (What to specify?), and not activities (How to specify?) and AM4RRE contains to this end the legal viewpoint and its relationships with engineering viewpoints (which is the core of AMDiRE) to assure coordination between the different viewpoints. In AM4RRE, coordination is achieved by specifying the information each viewpoint requires for regulatory requirements engineering and by exchanging this information among roles to support their viewpoint-specific objectives and the overarching goal of effective regulatory compliance.

The components of our artefact-centric approach are as follows:
\begin{enumerate}
    \item[C1] Role model (answering the question \textit{Who specifies?}) describes responsibilities and qualifications of the involved roles---C1.1 legal expert, C1.2 domain expert, C1.3 requirements engineer, C1.4 software architect.
    \item[C2] Goal model (\textit{Which goals are pursued?}) specifies the goals that each role is striving to achieve in regulatory RE and includes C2.1 legal expert goal model, C2.2 manager goal model, C2.3 requirements engineering goals model, C2.4 software architect goal model. 
    \item[C3] Project context model (\textit{Why?}) includes auxiliary project context information. We include C3.1 description and prioritization of challenges to be addressed in regulatory RE and C3.2 RE activities as components of such conceptual model.
    \item[C4] Milestone model (\textit{When?}) establishes the sequence of content items' specification.
    \item[C5] Content model (\textit{What is specified?}) defines the content of the specifications conducted by each role involved in regulatory RE. These specifications are: C5.1 legal specification, C5.2 software context specification, C5.3 requirements specification, and C5.4 system specification. Each of these specifications (often referred to as artefacts) is composed of content items (smallest semantically coherent aggregation of information having same purpose and meaning), which in turn are composed of concepts (smallest element of viewpoint information capturing phenomena and structuring the meaning of content item).
\end{enumerate}

Components of the AM4RRE (C1-5) belonging to one single role constitute one of four corresponding viewpoints (V1. legal, V2. business, V3. requirements, and V4. architecture viewpoints. The pivot of coordination in AM4RRE is the artefact model (C5) describing the content of specifications that roles are responsible for. The legal specification and our approach for defining it play a foundational role in identifying the information required from other roles and drive the specification of such information.

Before AM4RRE can be used in practice for defining the required information, it requires \textbf{tailoring}, i.e. the systematic adaptation of the AM4RRE to specific regulation, project, and/or goals in regulatory RE.
Tailoring can be applied at organizational, project or at both levels (depending on organizational setup) in one of the following ways: [T1] regulation-specific tailoring (mandatory), [T2] project content-specific tailoring (mandatory), [T3] goal-driven tailoring (optional).

T1. Regulation-specific tailoring is conducted to identify the information that needs to be specified in the content model (C5.2) to address specific regulation across different viewpoints and includes the following steps. T1.1: annotating the text of regulation to identify the instances of the legal concepts underlying the content model of legal viewpoint (e.g., concept of legal subject)). T1.2: all the required attributes of such concepts are identified (for example, age as a property of legal subject). For that, all the attributes for each content item in AM4RRE are first checked; however, an extension of AM4RRE with new concepts can be required. T1.3: tailored content model for legal specification is developed containing only content items, concepts and attributes identified in the text of regulation. If regulation-specific tailoring is conducted at the level of the organization, annotated text along with the content model represent an organization-wide interpretation of the regulation. It can be then complemented on the level of projects for complete instantiation of the model and interpretation of regulation.

T2. The purpose of content-specific tailoring is to establish relations between content specified in the legal specification and engineering specifications (context, requirements, and system specifications) and when required embed the required content to engineering specifications. It is conducted in the following steps. T2.1: for each legal concept in the legal viewpoint, we establish concepts in engineering specifications (e.g., concept \textit{age} of the content item <<legal subject>> is related to concept \textit{age} of the content item <<stakeholder>> of context specification). T2.2: it is checked that both concepts in the established relationship are defined in compatible ways (e.g., <<stakeholder>> in context specification satisfies <<data subject>> in legal specification or that \textit{age} is defined consistently), in case corresponding concept (e.g., \textit{age}) is absent in engineering specification it needs to be added to the most suitable content item in engineering specifications (e.g., if \textit{age} it should be added to content item <<stakeholder>>); in this case, the definition of the content element should be given on the basis of definition of \textit{age} in legal content item. This way, duplications can be avoided as content element in legal specification should be related to one content elements in one of engineering specifications. This contrasts with other approaches that may duplicate same requirements at requirements and system specification layers of abstraction. After this tailoring the established relations are used to guide the roles towards specific artefacts, content items and data elements in the process of regulatory RE.

T3. Goal-driven tailoring is optional, but can be used to better scope regulatory RE. For this, the goals relevant for organization are selected using the goal model. After that, relationships between goals and content items in content model are used to find out which content items need to be specified to achieve the corresponding goals (for example, if the goal of effective management of non-compliance is important <<sanction>> content item should be specified).

\section{Future Work}
The final steps in this doctoral study are the finalization of AM4RRE and its validation with practitioners. Validation, however, requires appropriate scoping because of the extensiveness of the model that can require longitudinal validations and, simultaneously, reluctance of practitioners to be involved in the studies related to regulatory compliance because of the sensitivity of the topic. In connection to this, we are considering two options for the model operationalization which are checklists and templates. We are also considering two options for the model validation. The first one consists of the involvement of experienced professionals in privacy engineering to conduct a walkthrough of all four viewpoints constituting AM4RRE and complementing the walkthrough with an experiment where participants specify requirements using an exemplary case, viewpoint specifications, and corresponding templates and checklists. The second option is to involve professionals in an experiment in which they will be assigned with one of the four roles and a corresponding viewpoint to address an exemplary case using the provided templates or checklist and specify requirements. In either case, the application of the experiment will require tailoring to account for complexity of validation involving four participants representing four different viewpoints (in an in-vivo setting). By presenting this work and the options for the validation, we hope to yield a fruitful discussion and receive feedback and suggestions from the community about the synthesis of the final model on the basis of the our previous research results and the operationalization and validation of AM4RRE.

\section*{Declaration on Generative AI}
During the preparation of this work, the author used Grammarly, ChatGPT in order to: Grammar and spelling check, Paraphrase and reword. After using this tool/service, the author reviewed and edited the content as needed and takes full responsibility for the publication’s content.

\bibliography{bibliography}

\end{document}